\documentclass[12pt]{article}

\usepackage[intlimits]{amsmath} 
\usepackage{amsthm,mathrsfs}    
\usepackage[comma,authoryear]{natbib}
\usepackage{url}
\numberwithin{equation}{section}
\usepackage{amssymb,amsmath}
\usepackage{subfigure}
\usepackage{bm}
\usepackage{booktabs} 
\usepackage{graphicx,graphics,epsfig}
\usepackage{float}

\theoremstyle{plain}
\newtheorem{thm}{Theorem}

\newcommand{\bthm}{\begin{thm}}
\newcommand{\ethm}{\end{thm}}

\newcommand{\bpf}{\begin{proof}}
\newcommand{\epf}{\end{proof}}

\theoremstyle{definition}

\newtheorem{exmp}[thm]{Example}

\usepackage[a4paper,text={16.5cm,25cm},centering]{geometry}
\usepackage[compact]{titlesec}
\newcommand{\bib}{\bibliography{ref-bib}\bibliographystyle{ims}}
\usepackage{deep-macro}

\begin{document}
\begin{center}
{\Large {\bf CDfdr: A Comparison Density Approach to Local False Discovery Rate Estimation}}
\\[.2in]
Subhadeep Mukhopadhyay\\
Temple University, Philadelphia, PA, USA\\[.4in]

{\bf ABSTRACT}\\
\end{center}

This paper defines a new class of methods for local false discovery rate (fdr) estimation based on two concepts: comparison density \citep{parzen83a,parzen99,parzen04b} and pre-flattening smoothing \citep{parzen79}. A simple, non-parametrically guided estimator is proposed, which allows researchers to efficiently model the tails of the ratio of two densities $f$ (marginal density) and $f_0$ (null density) directly in a single step, by properly expressing them in the quantile domain. Specific consideration is given to build a flexible, yet parsimonious parametric model that can be easily interpreted and implemented. We have also shown how almost all of the existing local fdr methods can be viewed as proposing various model specification for comparison density - unifies
the vast literature of false discovery methods under one concept and notation. Detailed discussion on estimation, inference, model selection and goodness-of-fit is given. Application to a variety of real and simulated data sets show promise. We end with some open problems.
\vspace*{.35in}

\noindent\textsc{\textbf{Broader Significance}}:  \cite{efron01} proposed empirical Bayes formulation of the
frequentist Benjamini and Hochberg’s False Discovery Rate method \citep{BH95}. This article attempts to unify ‘the two cultures’ using concepts of comparison density and distribution function, which could have far reaching consequences and impact.

\vspace*{.35in}

\noindent\textsc{\textbf{Foundation}}: The present work is an example of successful application of recently developed theory on `United Nonparametric Data Science' by \cite{deepthesis,D12b,D13a,D13USA,D12e}.

\vspace*{.4in}


\newpage

%

\section{Introduction}
 This paper introduces a new class of smooth nonparametric models to characterize the local false discovery rates (fdr) combining two concepts: comparison density and pre-flattened smoothing.

A very important goal of modern large-scale inference problems can often be framed as a signal-noise separation problem which starts with test statistics $T_1,\ldots,T_N$ with large $N$ (say, $N=10^4$). Local fdr, introduced in \cite{efron04}, provides an elegant framework for this purpose. The local fdr is defined as the conditional probability of a case being null or noise given $T=t$,
\beq
\fdr(t)\,=\, \Pr(\rm{null} \mid T=t)\,=\, \Pr(Y=0) \,\dfrac{f(t; T \mid Y=0)}{f(t;T)} \,=\, \pi_0 \,\dfrac{f_0(t)}{f(t)},
\eeq
where $Y$ is an $1/0$ indicator variable denoting the non-null and null cases respectively \footnote{Although the problem seems similar to two sample classification problem, the learning set does not contain any label information. For this reason we prefer to call it a one sample detection problem.}. We declare the cases interesting for which $\hfdr(t_n)$ is small, say $\hfdr(t_n) \le 0.1$ or $\hfdr(t_n) \le 0.2$. Clearly, accurate inference depends critically on efficient estimation of local fdr. Existing approaches estimates \textit{separately} $\hpi_0$, $\hf_0(t)$ and $\hf(t)$ to get an estimate of local fdr into expression (1.1) by
\beq
\hfdr(t)\,=\, \hPr(\rm{null} \mid T=t)\,=\, \hpi_0 \,\dfrac{\hf_0(t)}{\hf(t)}.
\eeq
It is no surprise, therefore that there has been an enormous amount of research done simply by plugging different density estimators in expression (1.2), e.g., Parzen-kernel density smoothing \citep{kerfdr}, normal mixture model \citep{Om10}, exponential family density \citep{efron04,efron08}, Bernstein density \citep{guan}, modified Grenander density \citep{strimmer} and many others.
It might well be the case as suggested by Benjamini \citep[p. 26]{benj08} ``the tools developed along with the approach may have reached the stage where it is unlikely that  further polishing of same tools will be of much help.'' This immediately raises two questions:

\vskip.4em
 (i) whether or not there is any scope of further improvement.
\vskip.25em
 (ii) if so, how can we advance current state-of-the-art to the next level ?
\vskip.4em
This paper proposes a new class of models called Comparison Density based False Discovery Rate (CDfdr) and suggests a data-driven adaptive estimation procedure for building such a model.  There are various attractive features of this alternative modeling strategy which make it suitable for wide class of data from diverse applications like bioinformatics, particle physics, astronomy, neuroimaging and so on. The main motivation of our methodology comes from the following two considerations:

\vskip.4em
 ({\bf A}) classical local fdr methods require an estimation of $f$ which create an additional obstacle in estimating local fdr without imposing any further assumptions on the rate of tail-decay. The question is, can we altogether bypass the problem of estimating marginal density $f$ ? This might alleviate the problem of higher variability by reducing the number of parameters to estimate. This is ``even more pronounced in the far
tails which is usually most important for large-scale screening'' \citep[p.23]{benj08}. Overall, one might expect more robust and data-analytic (less model dependent) estimates if we can avoid estimation of $f$.

\vskip.35em

 ({\bf B}) the second one is a more fundamental and crucial issue which comes from recognizing that estimating directly $f_0(t)/f(t)$ is much more efficient and straightforward than estimating them separately and then taking the ratio $\hf_0(t)/\hf(t)$. It is worth asking whether we can estimate local fdr (1.1) directly in one-step rather than using two-step approach; thus ensuring more stability and computational gain. There has not been any attempts so far in this particular direction, perhaps because, there is no standard tool available to accomplish this.
\vskip.4em

The main challenge is to develop a flexible yet simple algorithm for modeling tails of the ratio of two densities $f_0(t)$ and $f(t)$.
Our quantile based approach attempts to address this important applied problem, bringing new tools and concepts for large-scale discovery problems. It is demonstrated that CDfdr achieves parsimony and improves over leading methods in terms of accuracy of estimation (specially the tail region which is the main deciding factor). Other added advantage of the CDfdr algorithm are its  ``interpretability'' and that it is easy to implement. New motivation of multiple hypothesis testing problems from comparison density perspective is also given. The paper is written in a style which is highly applicable towards the culture of ``vigorous theory and methods for translational research''.

\vskip.3em

The rest of the paper is organized as follows: In Section 2 we give a brief description of the main idea and connect it with local fdr.  Section 3 deals with the estimation part. The concept of ``smooth'' p-value is introduced using the beta-preflattened transformation. In addition, we describe in detail the CDfdr algorithm. In Section 4 we illustrate our approach using prostate cancer data. Section 5 presents two simulation studies. Summary, conclusion and future direction are presented in Section 6.

\section{Model}

The main purpose of this section is to introduce a new tool for multiple testing problem using the concept of comparison density and to connect it with local false discovery rate.
\subsection{Comparison Density: Functional Inference Approach Towards Multiple Hypothesis Testing}
We will start by defining comparison density, the most important conceptual tool in our analysis. For continuous $F$ and $G$, the comparison density is defined as follows:
\beq
d(u;F,G)\,=\, \dfrac{g(F^{-1}(u))}{f(F^{-1}(u))}, \quad 0 < u<1.
\eeq
The concept of comparison density can be motivated from various angles. Here we will discuss how comparison density naturally arises in the context of simple hypothesis testing and goodness of fit. Consider $T_1,\ldots T_N$ to be a random sample from continuous $F$. The objective is to test $F=F_0$. This problem can be converted into \textit{testing uniformity} of $U=F_0(T)$, whose distribution function is $F(F_0^{-1}(u)):= D(u;F_0,F)$ and quantile function $F_0(F^{-1}(u)) = D(u;F,F_0)$. In this set up, the comparison density $d(u;F_0,F)$ is the density of $U$. Given this new formulation, the testing problem $F=F_0$ can now be recast as testing $D(u;F_0,F)=u$ for all $0<u<1$ or equivalently:
\beq
d(u;F_0,F)\,=\,1, ~\text{for all } ~ 0<u<1.
\eeq
The notion of $d(u;F_0,F)$ helps to transform the hypothesis testing problem into a ``functional statistical inference'' problem \citep{parzen83a}, which act as a liaison between comparison density and local fdr. To better understand the implication for multiple hypothesis testing, note that the collection of $u$'s for which $d(u;F_0,F)$ \textit{substantially} deviates from $\rm{Uniform}[0,1]$, are precisely the non-null candidates that we are searching.\textit{ This notion of functional approach using comparison density enables us not only to test the hypothesis but also to detect the interesting cases; thus acting as an useful tool for large-scale hypothesis testing problems}. Two questions remain unanswered:
\vskip.1em
(i) what is the guarantee that this (heuristic) algorithmic approach will work ?
\vskip.14em
(ii) and secondly, how to decide on the threshold to detect substantial deviation of $\dhat(u;F_0,F)$ from uniformity.
\vskip.1em
In the next section, we will connect this idea of comparison density based multiple hypothesis testing with the local false discovery rate to answer these questions.

\subsection{Towards An Alternative Formulation:  Comparison Density based One-Step Approach}
In the previous sections, we have introduced the idea of comparison density and heuristically explained how it plays a crucial role in large-scale inference problems. Here we give an alternative representation of the local false discovery rate using the comparison density combining (1.1) and (2.1).

{\bf Proposition 1.}~~~~~~~~$\fdr(t) ~:= ~\Pr\{ \rm{null} \mid T=t\}~ =~  \dfrac{\pi_0}{d( F_0(t); F_0, F )}.$
\vskip.35em

Proposition 1 gives an alternative way of modeling local fdr via comparison density where different comparison density estimation methods will generate different model specifications for local fdr. Proposition 1 also allows us to transform the problem of local fdr estimation as a special type of density estimation problem. We call this general class of models the comparison density based local false discovery rate (CDfdr). The main benefit of introducing comparison density follows from the observation that comparison density is nothing more than density of the p-values $F_0(t)$, which can be estimated in a single step that is formalized into the following result.

{\bf Proposition 2.}~~~~~~~~$d(u;F_0,F)\,=\, f(F_0^{-1}(u))/f_0(F_0^{-1}(u))\,=\, \rm{density\, of\, } U.$

\bpf
The distrbution function of $U$ is given by \beq \Pr(F_0(T) \le u) = \Pr(T \le F_0^{-1}(u)) = F (F_0^{-1}(u)),\eeq
which implies that $ \left( F(F_0^{-1}(u)) \right)' = f(F_0^{-1}(u))/f_0(F_0^{-1}(u)) = d(u;F_0,F)$.
\epf
Proposition 1 and 2 are extremely useful as we can now hope to achieve the goal of estimating local fdr directly in a single step, without requiring estimation or specification of the marginal density $f(t)$. It is important to realize that one can work with either the original test-statistics (distribution domain) or p-values (quantile domain). \textit{But one can not avoid the problem of two-step estimation in distribution domain}. Whereas quantile domain transformation (p-values) tackle the estimation directly in one-step using comparison density, which precisely the main message of Proposition 1 and 2. Also note that conventional threshold for reporting the interesting cases $\widehat \fdr(t_i) < .2$ \citep{efron04,efron08} is equivalent
to (using 2.3 and assuming $\pi_0 \approx 1$)
\beq \dhat(F_0(t_i);F_0,F) > 5.\eeq
Expression (2.4) not only justifies the arguments made at the end of Section  2.1, but also establishes a judicious choice of threshold by connecting it with local fdr concept.

\subsection{New Challenge: Tail-Modeling}
The previous section suggests the following simple strategy for estimating fdr via CDfdr algorithm.
\beq \text{Test statistics} ~\rarrow~   \text{P-values}   ~\rarrow~    \text{Density of p-values}    ~\rarrow~   \text{CDfdr}.       \eeq
\vspace{-.1em}
Let's illustrate it using the following example:
\vspace{-.1em}
\begin{exmp}[golub gene expression data, \cite{golub99}]
$N=7129$ gene expressions on Leukemia cancer study; comparing $n_1=27$  acute lymphoblastic leukemia (ALL) and $n_2=11$ acute myeloid leukemia (AML) tumor samples to identify differentially expressed genes (available in Bioconductor R package \texttt{golubEsets}); p-values based on two-sample t-test analysis are shown in the panel A of Fig 1.
\end{exmp}
We quickly realize that direct application of the algorithmic steps (2.5) using conventional density estimation techniques for golub p-values encounter obstacles. The main challenge comes from modeling the sharp narrow peak near the boundary $0$, indicating the presence of the signal(sparse), as illustrate in Fig 1A.

\vskip.3em
There is an impressive list of techniques available for estimating comparison density or the density of the p-values, kernel density smoothing, regression based density estimators, exponential series density estimators and many others. For a comprehensive list see \cite{rd99}. Although all of these methods enjoy excellent theoretical properties, their utility for the situation at hand is questionable. It is well-known that kernel density estimation suffers from the ``boundary effect'' as illustrated in Fig 1B. Regression based density estimators via smoothing splines or local polynomials, are known to have have a larger variance near the boundaries \citep{Thas10}. One would expect the exponential series density estimator to be heavily parametrized to capture the tail, which may lead to undesirable spurious bumps. To tackle this highly dynamic non-standard density estimation problem, we propose a specially designed procedure - particularly suitable for large-scale signal detection problems. At the heart of this new method lies the  concept of \textit{pre-flattening or pre-whitening}, as suggested in \cite{parzen79}.

\vskip.15em
\begin{figure*}[!htb]
 \centering
 \includegraphics[height=.7\textheight,width=\textwidth,keepaspectratio,trim=.5cm .5cm .5cm .5cm]{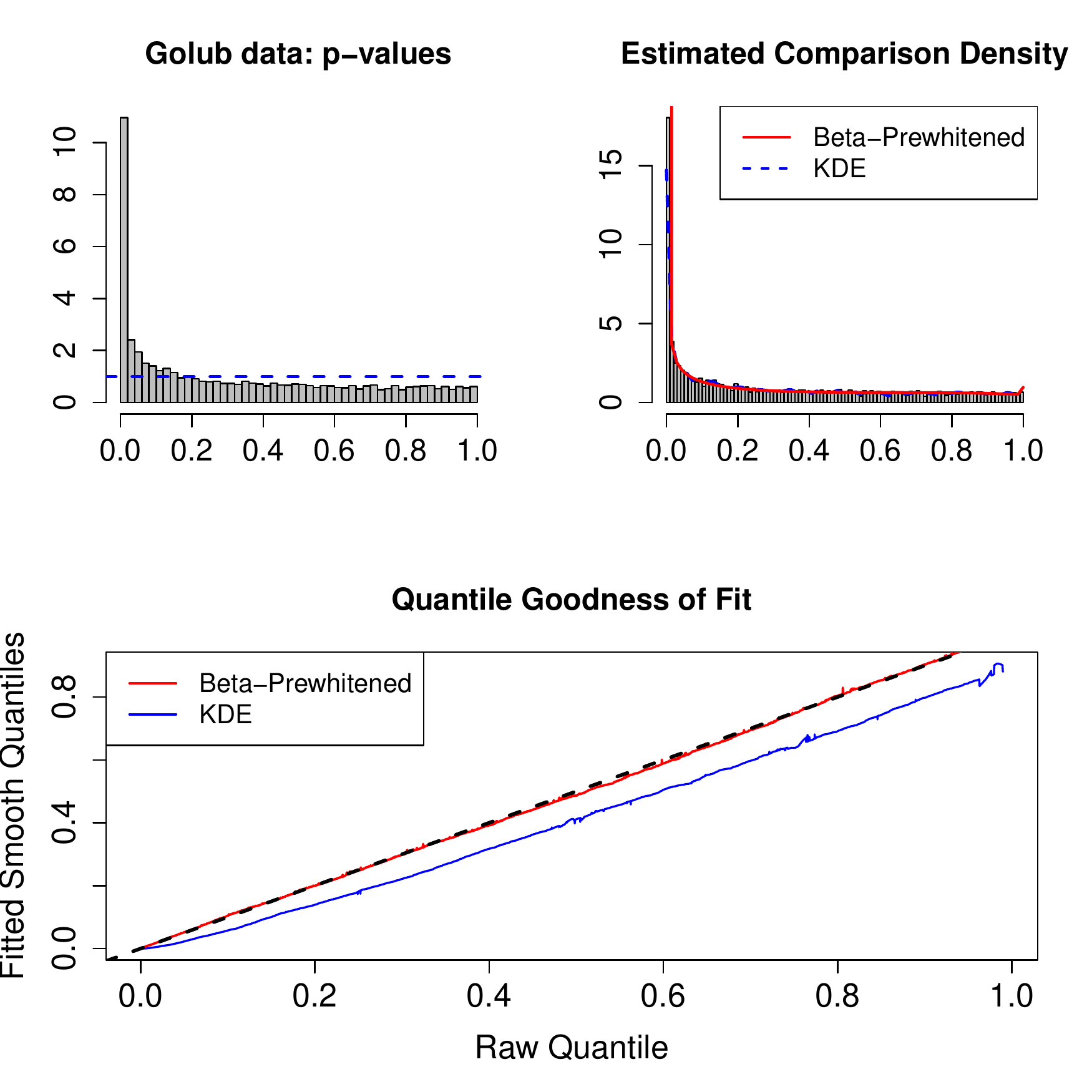} \\
\vspace{-.5em}
\caption{(A) Histogram of the $7129$ p-values of golub gene expression data using two sample t-test. (B) Fitting of kernel density smoother and beta-preflattened density estimate (introduced in Section 3). (C) Goodness of fit to compare the fitting. We compared the raw quantile with the smooth quantile function from the two competing models. Unarguably beta-flattening gives much better fit even in the extreme tail.}
\end{figure*}
\vskip.15em

Overall, it is not difficult to propose new density estimation techniques to fit the data such as Fig 1A but it is less easy to come up with a parsimonious parametric model which fits the data well and is easy to interpret. The most stunning fact about our beta-flattening approach (that we will elaborate in the next section) is that, \textit{it required only three parameters to model the golub pvalues satisfactory well including the tail region} !

\section{Estimation}
We aim to develop a bonafide parametric model for CDfdr ensuring sparsity, smoothness and flexibility. Our proposal consists of two main steps:
\vspace{-.5em}
\begin{itemize}
 \item convert ``spiky'' p-values to ``smooth'' p-values via the preflattening technique that we will describe shortly. The novelty is in the choice of pre-flattening function which we choose as beta density to efficiently capture  the rapidly changing tails.
 \item Estimate smooth p-values using adaptive orthogonal series density estimator.
\end{itemize}
\vspace{-.5em}
This technique has a unique ability to ``decouple'' the density estimation problem into two separate modeling problems: the tail part and the central part of the distribution.

\begin{figure*}[!htb]
 \centering
 \includegraphics[height=.41\textheight,width=\textwidth,trim=.3cm .5cm 0cm 0cm, clip=true]{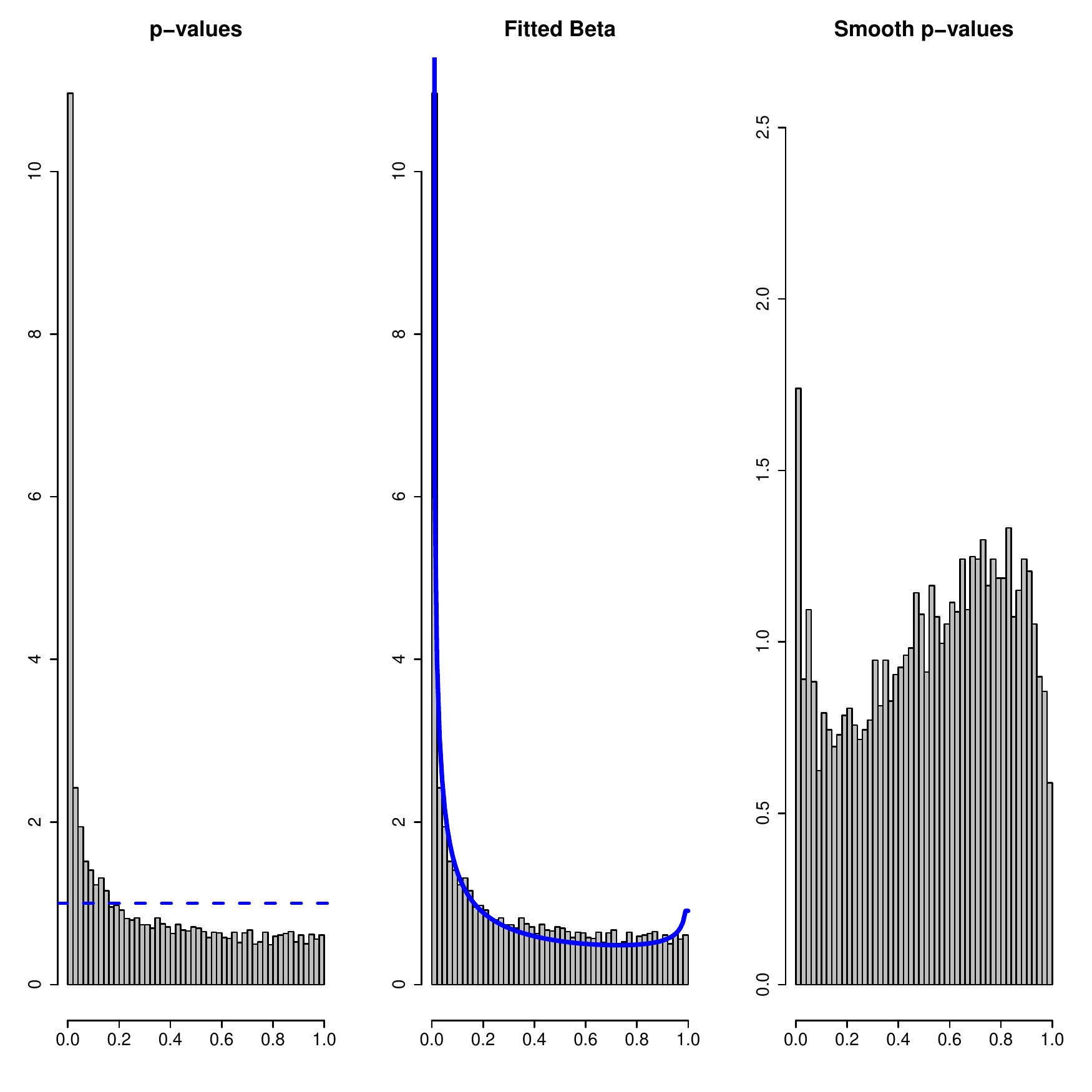} \\
\vspace{-.5em}
\caption{Converting p-values to smooth p-values for the golub data. This technique reduces the dynamic range of the original p-values by more than a factor of $10$.}
\end{figure*}
\vspace{-1em}

\subsection{Beta-Preflattening and ``Smooth'' P-values}
We have already recognized the difficulty of directly estimating the density of p-values. The tool described in this section will allow us to transform this difficult problem into a seemingly trivial one. Key idea is to decompose the comparison density into two parts,
\beq d(u;F_0,F)\,=\,f_{\rm{B}}(u;\,\al,\be)\, d\big( F_{\rm{B}}(u;\al,\be);\, F_{\rm{B}},F     \big), \quad 0<u<1 \eeq
where $f_{\rm{B}}$ denotes the beta density and $F_{\rm{B}}$ denotes the beta distribution function. Here the beta density with parameters alpha and beta $f_{\rm{B}}(u;\,\al,\be)$ act as the pre-flattening function. We define the quantity $F_{\rm{B}}(u;\al,\be)$ as the ``smooth'' p-value.
\vskip.3em
The equation (3.1) can be interpreted from an algorithmic point of view as:
\beq \text{Density of the original p-values} \,=\,\text{ Fitted beta } \, \times \, \text{Density of the smooth p-values}.\eeq
Fig 2 shows the implementation by first fitting a $\rm{Beta}(\hat \al=.32,\, \hat \be=.75)$ to the p-values and then generating the smooth p-values $v\,=\,F_{\rm{B}}(u;\,\hat \al=.32, \hat \be=.75)$. In the next section we will describe a simple procedure to estimate the density of $v$. The point worth emphasizing here is that even though the density estimation  of $d(u;F_0,F)$ is a quite challenging task (left of Fig 2), pre-whitening by beta density and estimating the density of smooth p-values (right of Fig 2) is an incredibly simple and a straightforward exercise.

\begin{figure*}[!ht]
 \centering
\includegraphics[height=.5\textheight,width=\textwidth,trim=.4cm .5cm 0cm 0cm, clip=true]{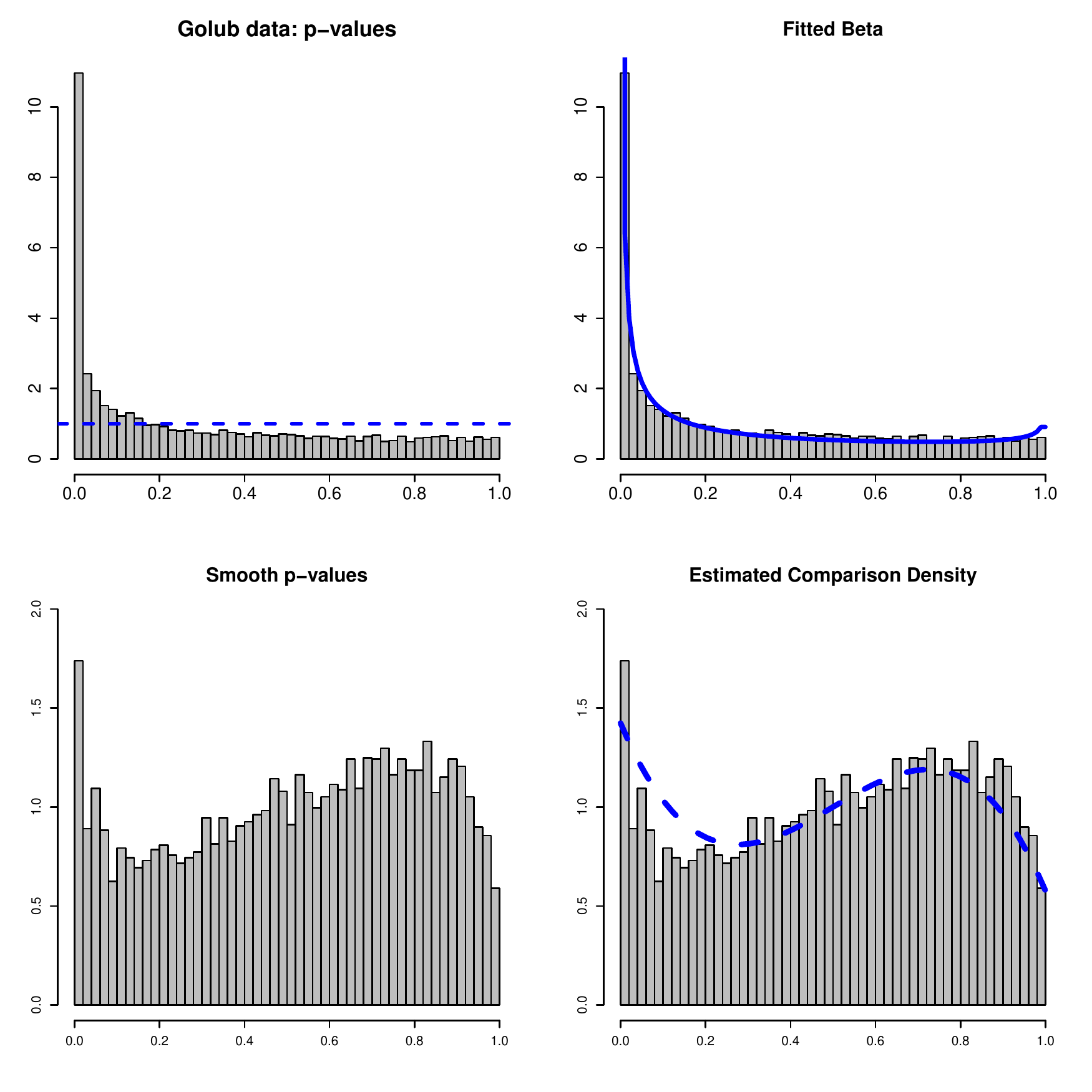}
\vspace{-.5em}
\caption{Mechanism of beta-preflattened density estimation; \textbf{I}. Fit beta density. For golub data it is $\rm{Beta}(\hat \al=.32,\, \hat \be=.75)$; \textbf{II}. Generate smooth p-values; for golub data it is $\,F_{\rm{B}}(u;\,\hat \al=.32, \hat \be=.75)$; finally, \textbf{III}. Estimate the density of smooth p-values by adaptive orthogonal density estimator; for golub data $\dhat(v;\, F_{\rm{B}},F ) = 1\, -\, .16 S_3(u)$, where $S_3$ is the third orthonormal Legendre polynomial on $[0,1]$.}
\end{figure*}

\subsection{Estimating density of ``Smooth'' P-values}
Note that the density of smooth p-values is a well-behaved bounded function (see Fig. 2), which can now be modeled using conventional density estimation techniques. Here we will use an adaptive orthogonal series density estimator. We choose our basis functions $S_1,S_2,\ldots $ as shifted orthonormal Legendre polynomials over $[0,1]$. Under this framework, the smooth non-parametric model for comparison density has the form:
\beq
d(v;F_{\rm{B}},F)\,=\,1\,+\,\sum_{j=1}^M \te_j \, S_j(v), \quad 0<v<1.
\eeq
The preference for a such model solely comes from the consideration of simplicity and ease of estimation which will be clear soon. However, the reader is free to choose any other density estimators for $v\,=\,F_{\rm{B}}(u;\,\hat \al, \hat \be)$.

Note that the score coefficients can be quickly estimated from the p-values $u_1,\ldots,u_N$ by
\beq
\widetilde \te_j\,  \larrow \, N^{-1}\sum_{i=1}^N S_j\big( F_{\rm{B}}(u_i;\,\hat \al, \hat \be) \big),
\eeq
which follows from verifying the following fact,
\vskip.22em

{\bf Proposition 3.}~~~~~~~~$\te_j\,=\, \Ex\big[\, S_j \big(F_{\rm{B}}(U;\,\al, \be) \big)\, \big]$.
\bpf
Note that,
\beq
\te_j\,=\, \int_0^1 \dfrac{f(F_0^{-1}(u))}{f_0(F_0^{-1}(u))}\, S_j(v) \dd v \,=\, \Ex\big[ S_j(F_0(Y)) ; F     \big],
\eeq
where the we get the second equality by substituting $F_0(y)=v$. By virtue of $F_0(Y)=V=F_{\rm{B}}(U;\,\al, \be)$ the result follows.
\epf

The data-driven sparse model for smooth p-values is given by
\beq
d(v; F_{\rm{B}},F)\,=\,1\,+\,\sum_j \widehat \te_j S_j(v), \quad 0<v<1.
\eeq
where the score coefficients $\widehat \te_j$ is defined as,
\beq \hte_j\,=\, \widetilde \te_j\, \ind_{\bigl\{ \widetilde \te_j^2 \,>\, 2 N^{-1} \log N \bigr\}} . \eeq
This reduces spurious oscillations and renders stability by adapting to the  underlying smoothness. For golub data the thresholded estimator of score coefficients only selects $\hte_3=-.16$ out of $M=6$ (3.3). From our experience, as distribution of $v$ is quite smooth (already pre-whitened) this choice of $M$ works in most cases.

It turns out that this data-adaptive orthogonal series density estimator has nice theoretical properties which are thoroughly discussed in \cite{ander80,ledwina94,efro99}. Also note that large $N$ (for golub data $N=7129$) makes all of these asymptotic analysis very much relevant and directly applicable.

\subsection{$d(u)$ Assisted New Density Estimation Technique}
The technique suggested in the previous section can be easily generalized as a certifiable density estimation tool by writing
\beq
f(x)\,=\,f_0(x)\, \dfrac{f(x)}{f_0(x)}\,=\,f_0(x)\, d\big( F_0(x)   ;\,F_0,F   \big).
\eeq
Any density can be represented as (3.7), where $f_0$ is our parametric pre-whittening function and $d(u)$ is the corresponding comparison density. An example would be skew normal where we first fit a normal and estimate the $f_0$ and in the next step we estimate non-parametrically the density of the rank-$F_0$ transform data $u=\widetilde F_0(x)$. Further details can be found in the papers by \cite{D12b,D12e,D13USA}.
\subsection{Estimating Proportion of True Null Hypothesis}
We now turn to the problem of estimating $\pi_0$, the true proportion of noise or null hypothesis which is an integral part of the definition of local fdr (Proposition 1). Recent works include \cite{storey03,lan05,jin07}. Here we develop a data-analytic algorithm which utilizes the beta-preflattened nonparametric comparison density estimator
\beq
\dhat(u;F_0,F)\,=\,f_{\rm{B}}(u;\,\widehat \al,\widehat \be)\, \dhat\big( F_{\rm{B}}(u;\widehat \al,\widehat \be);\, F_{\rm{B}},F     \big), \quad 0<u<1.
\eeq
We begin by stating the algorithm.

\vskip.3em
\textbf{Algorithm 1} [$\pi_0$ estimation by  Minimum Deviance Criteria (MDC)]
\vskip.2em
\textit{Step 1.} Define $\cU_{\la}=\{u: \dhat(u;F_0,F) < \la\}$; $|\cU_{\la}|=N_{\la}$. For each fixed $\la$ on a fine grid between $[1,3.5]$ repeat the following steps.
\begin{itemize}
 \item[(a)] Compute $\tilde \te_j^{\la} \,\larrow \, N_\la^{-1} \sum_{i=1}^{N_\la} S_j(u_i)$ which is the score coefficient for the following $L_2$ comparison density $ \tilde d_{\la}(u)\,=\, 1\,+\, \sum_{j=1}^M \tilde \te_j^{\la}\, S_j(u)$,  based on $\cU_{\la}$.
\item[(b)] Calculate the deviance statistic $D_{\la} \,\larrow \, \sum_{j=1}^M \big| \tilde \te_j^{\la} \big|^2$.
\end{itemize}
\textit{Step 2.} Display the deviance path $(\la, D_{\la})$ for $1 \le \la \le 3.5$ and set $\la^{*} \,\larrow \, \arg\min_{\la} D_{\la}$.
\vskip.2em
\textit{Step 3.} Output $\widehat \pi_0$, proportion of p-values satisfying $\dhat(u;F_0,F) < \la^{*}$.
\vskip.35em
The rationale behind the algorithm comes from the simple fact that under $H_0$ when all the cases are null the underlying comparison density should not deviate much from $\rm{Uniform}[0,1]$ (2.2). The statistic $D_{\la}$ quantifies the deviation of $\tilde d_{\la}(u)$ from uniformity.
Fig 4. illustrates this idea for Prostate cancer data sets (described in Section 4) where the shape of the smooth estimated comparison density clearly indicates the presence of signal in the two tails. The deviance path for $1 \le \la \le 3.5$ is shown in the right panel for $M=10$ which gives $\arg\min_{\la} D_{\la}=1.98$ and $\widehat p_0=0.971$.
At the point $\la^{*}=1.98$ the deviance statistics takes the minimum value which implies that $\tilde d_{\la=1.98}(u)$ is closest to the uniformity or in other words, the set $\cU_{\la=1.98}$ is most likely the null cases. The MDC can also be linked with the concept of signal to noise ratio.

\subsection{Estimating True Nonnull Density}
We can also estimate the density of the nonnull cases $f_1(t)$ using the estimated comparison density (3.8). Straightforward calculation shows
\beq
f_1(t)\,=\, \dfrac{1}{1-\pi_0} \left[  d\big( F_0(t);F_0,F \big) \,-\, \pi_0          \right]\, f_0(t).
\eeq
As is discussed in \cite{efron07}, the nonnull density and its different functionals play a vital role in power diagnostic. The equation (3.9) essentially says that we can reconstruct the nonnull density by appropriately weighting the null density. Fig 4 (A) helps to better understand the weighting scheme which is almost zero in the central region and significantly increases as we move in the tails allowing possible asymmetry, thus it learns the ``modal-sharpness'' in a data-analytic way.

\subsection{CDfdr Algorithm: Tree and Forest View}
We now combine all the ideas and develop a computationally simple implementation strategy which can efficiently handle large data sets. Our CDfdr algorithm for estimating local fdr can be described as follows:

\vskip.35em
\textbf{Algorithm 2} [CDfdr algorithm]
\vskip.35em
\textit{Step 1. Quantile transformation: } Perform rank-$F_0$ transformation to get the p-values
\beq u_i\, \larrow\, F_0(t_i), \quad i=1,\ldots, N.\eeq
$F_0$ could be either theoretical null or empirical null provided to the CDfdr algorithm by user.
\textit{Step 2. Fit Beta density: } Compute $\widehat \al$ and $\widehat \be$ mle from $u_1,\ldots,u_N$. Efficient implementation available in R package \texttt{fitdistrplus}.
\vskip.35em
\textit{Step 3. ``Smooth'' p-values: } Transform p-values $u_1,\ldots,u_N$ to smooth p-values $v_1,\ldots, v_N$ by
\beq v_i \larrow F_{\rm{B}}(u_i;\widehat \al,\widehat \be) , \quad i=1,\ldots, N.\eeq

\textit{Step 4. Pre-flattened comparison density estimation: } Estimate density of pre-flattened $v_1,\ldots, v_N$ using orthonormal shifted Legendre polynomials (Section 3.2)
\beq  \dhat(v; F_{\rm{B}},F)\,=\,1\,+\,\sum_j \widehat \te_j S_j(v), \quad 0<v<1.   \eeq
We estimate $\widehat \te_j$ by (3.4) and (3.6).
\vskip.35em
\textit{Step 5. $\pi_0$ estimation: } Use Minimum Deviance Criteria (MDC) (Algorithm 1).
\vskip.35em
\textit{Step 6. Output $\widehat \fdr$: } Finally we generate the smooth (non)parametric estimate of local fdr
\beq \widehat \fdr(t)\,\larrow \, \widehat \pi_o \left[ f_{\rm{B}}\big(F_0(t);\,\hat \al,\hat \be \big)\,\times\, \dhat \big( F_{\rm{B}}(F_0(t);\,\hat \al,\hat \be ) ;\, F_{\rm{B}},F   \big)     \right]^{-1} .\eeq

\begin{figure*}[!htb]
 \centering
 \includegraphics[height=.5\textheight,width=\textwidth]{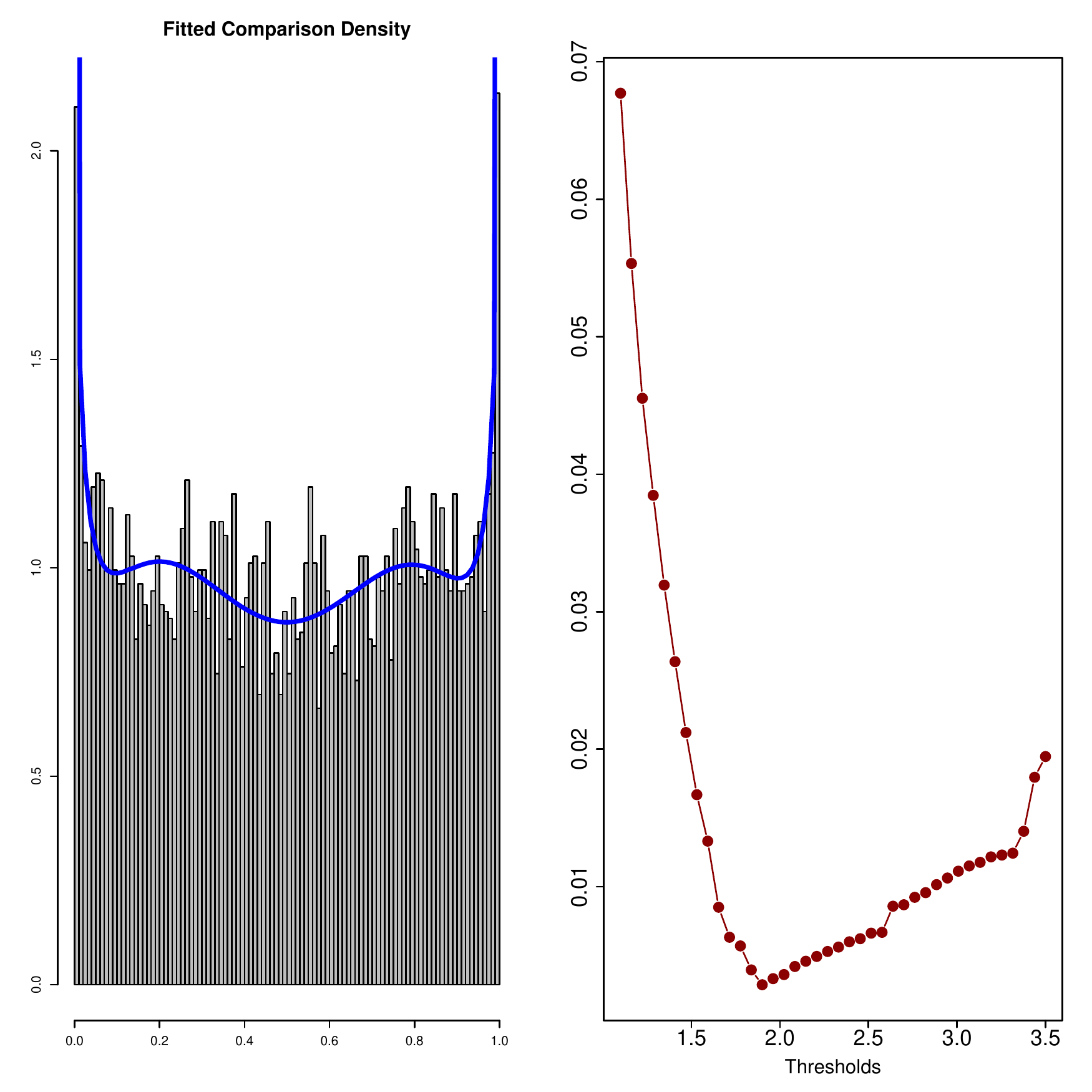} \\
\vspace{-.5em}
\caption{Estimated smooth nonparametric comparison density for Prostate cancer data and the deviance curve to determine $\widehat p_0$. }
\end{figure*}

\section{Real Data Application}
In this section we analyze the prostate cancer data \citep{prostate}. Our goal is to interpret each modeling step of CDfdr (Algorithm 2) and finally compare the result with the few leading methods.

The data consists of $102$ patient samples ($50$ labeled as normal and $52$ as prostate tumor samples) and $6033$ gene expression measurements. We aim to detect interesting genes which are differentially expressed in the two samples. For that purpose we compute the two sample t-test statistic $t_i$ for each gene and convert them into z-scale by $z_i~\leftarrow ~ \Phi^{-1}(\mathcal{T}_{100}(t_i))$, where $\mathcal{T}$ denotes the t-distribution function, shown in panel A of Fig 5. At the next step we fit the $\rm{Beta}(\widehat \al=.81, \widehat \be=.82)$ to the p-values to get the smooth p-values $ F_{\rm{B}}(u;\widehat \al=.81,\widehat \be=.82)$ shown in panel C of Fig 5; estimate the pre-whitened comparison density \beq \dhat(v; F_{\rm{B}},F)\,=\,1+ 0.057 S_6(v).\eeq
Fig 4 shows the final beta-preflattened smooth estimate of comparison density given by
\beq
\dhat(u;\Phi,F)\,=\, .68 \,\big[ 1\,+\, 0.057 S_6 \big( F_{\rm{B}}(u;\widehat \al=.81,\widehat \be=.82) \big)  \big]\,u^{-.19}\,(1-u)^{-.18},\quad 0<u<1.
\eeq


\begin{figure*}[thb]
 \centering
 \includegraphics[height=.45\textheight,width=\textwidth,trim=1cm 1.5cm 0cm 0cm, clip=true]{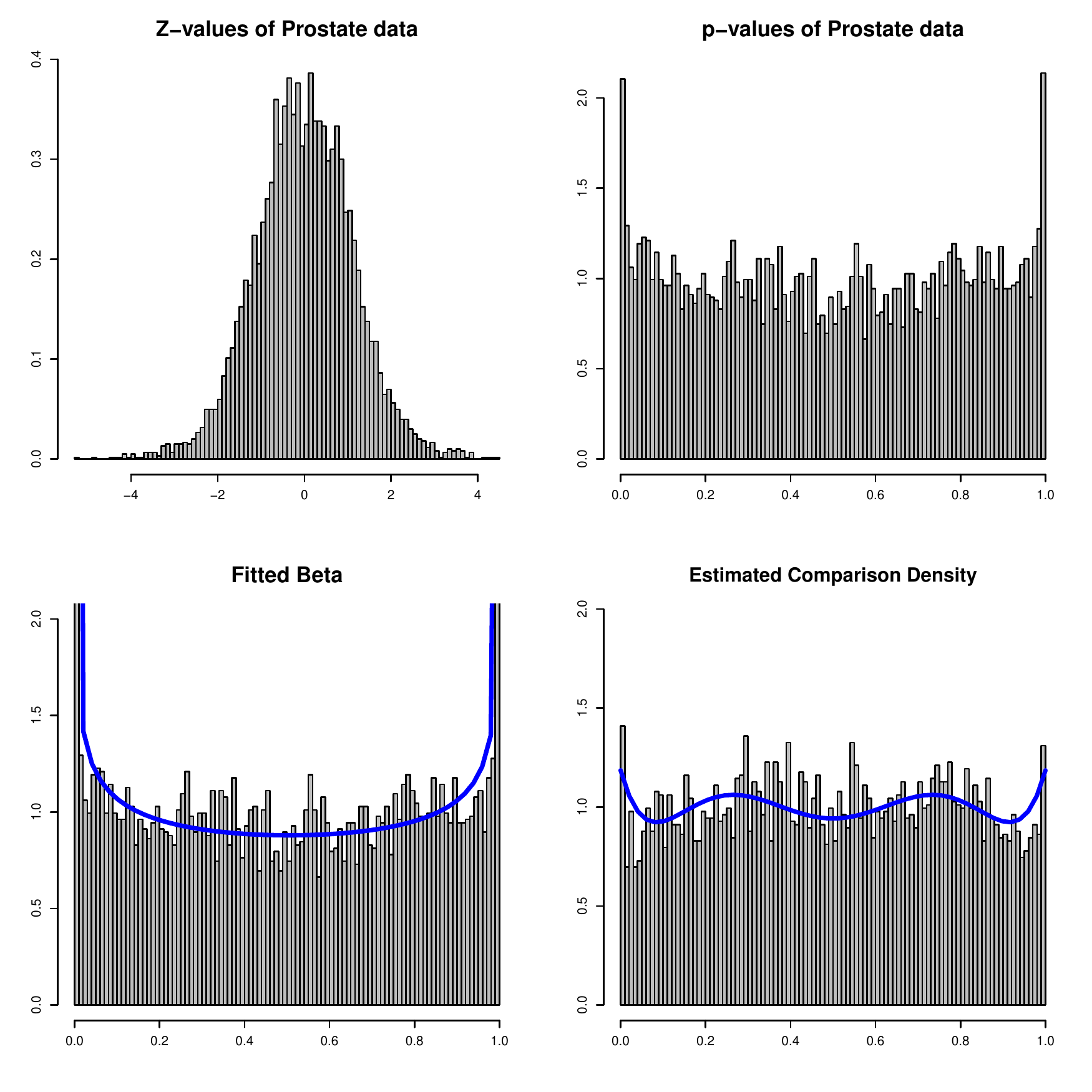} \\
\vspace{-.5em}
\caption{CDfdr in prostate data.$\hat d(v;F_{\rm{B}},F) = 1 + .057 S_6(v)$. }
\end{figure*}
\vskip.2em

which along with the Minimum Deviance Criteria (Algorithm 1, Fig 4B) gives $\widehat p_0=.971$. Consequently we can now estimate the local fdr using the representation (3.12).

\vskip.3em
We compare our result with Locfdr \citep{efron08} and Mixfdr \citep{Om10} that estimates the local fdr (1.1) by separately estimating the numerator $\hf_0$ and the denominator $\hf$. Naturally there are many variant available for these two methods depending on the way they estimate null and marginal density. We have used the R package \texttt{locfdr} and \texttt{mixfdr} for implementation purpose. Locfdr estimates pool destiny $f$ using splines. Methods for estimating null: (a) theoretical ($\cN(0,1)$); (b) maximum likelihood (MLE); (c) central matching (CM); (d) split-normal (SN). Mixfdr implement $J$ group normal mixture model for $f$. Estimation of empirical null involves putting Dirichlet prior on mixing proportion. We have used the default choice of $J$ and Dirichlet parameter $P$ throughout.
\vskip.25em
We First note two overall patterns from the result summarized in Fig 6.
\begin{figure*}[!thb]
 \centering
 \includegraphics[height=.5\textheight,width=\textwidth,keepaspectratio,trim=1cm 1.5cm 0cm 0cm, clip=true]{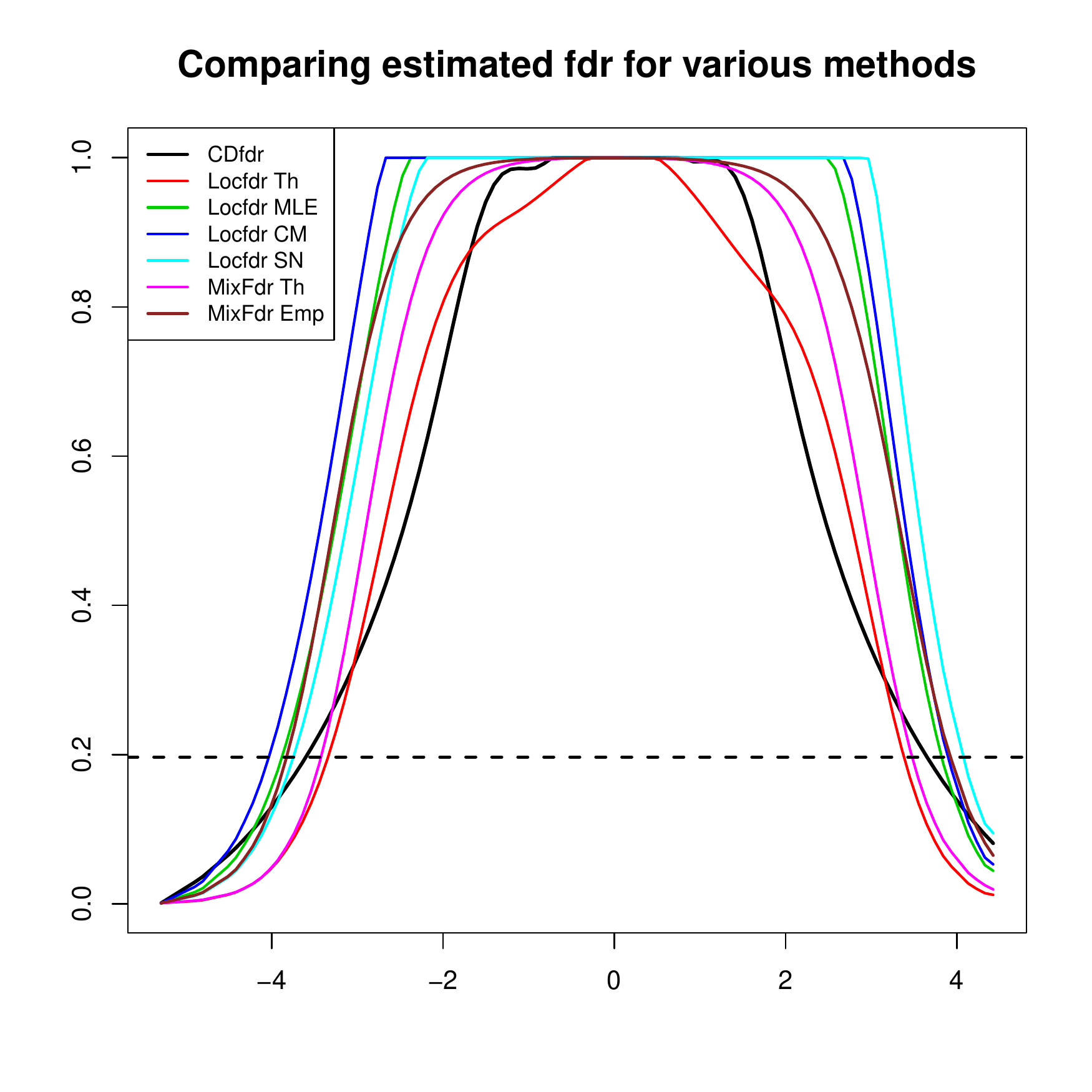} \\
\vspace{-.5em}
\caption{Estimated local fdr is shown for seven different methods for Prostate cancer data.}
\end{figure*}
in the estimated fdr curve for Locfdr and Mixfdr based on whether $f_0$ is theoretical null or empirical null. Interestingly, the CDfdr estimate is close to the empirical null estimates in the tail-regions and slowly matches with the theoretical null model in the central region, albeit CDfdr is implemented using theoretical null $\cN(0,1)$. This is further reflected in the Table 1. Number of non-null genes identified using CDfdr matches with the Locfdr-SN and Mixfdr-Emp method. Further, estimates of the proportion of true null matches with the estimates from Mixfdr-Th. Note that the estimates of $\pi_0$ from Locfdr-CM and Locfdr-SN are unrealistic as they exceed $1$. There is no doubt that the ultimate inference highly sensitive to the choice of null for Locfdr and Mixfdr. A small change in the null distribution has a massive impact on the inference. For example, using $\cN(-.003,1.087)$ (Locfdr-MLE) instead $\cN(0,1)$ reduce the number of discoveries more than $50\%$. We believe, the main reason for this instability has its root in the \textit{two-step} estimation process. Recall that, we tackle this problem by estimating the ratio directly via beta-preflattened trick, and by virtue of that one might expect more  reliable and reproducible inference from CDfdr algorithm.
\vskip.25em
As it is evident from Fig 6. the tails of the estimated fdr curves play a crucial role for separating signal from noise ($ \widehat \fdr < .2$). It might be more appropriate to focus on the quality of ``tail-modeling'' instead of considering the entire curve. We carry out this in the next section where we carefully quantify the estimation accuracy specially in the tails.

\begin{table}[!ht]
\setlength{\tabcolsep}{12pt}
\caption{\textit{Local fdr threshold $.2$ is used for detecting non-null cases. `$\leftrightarrow$' symbol is used to describe how many discoveries are on the left and right side. In the context of Prostate cancer data, how many under-expressed and how many over-expressed genes are interesting.}} 
\vskip1.5em
\centering 
\begin{tabular}{c c c c} 
\hline\hline 
Methods & \# discovery & $\widehat \pi_0$ \\ [0.5ex] 
\hline 
CDfdr & ${\bf 17}$  \,$(13 \,\leftrightarrow \, 4)$ & ${\bf 0.971}$\\
Locfdr-Th & $54$  \,$(27 \,\leftrightarrow \, 27)$ & $0.961$\\
Locfdr-MLE & $19$  \,$(12 \,\leftrightarrow \, 7)$ & $0.998$\\
Locfdr-CM & $13$  \,$(9 \,\leftrightarrow \, 4)$ & $1.015$\\
Locfdr-SN & ${\bf 17}$  \,$(13 \,\leftrightarrow \, 4)$ & $1.009$\\
Mixfdr-Th & $49$  \,$(26 \,\leftrightarrow \, 23)$ & ${\bf 0.971}$\\
Mixfdr-Emp & ${\bf 17}$  \,$(13 \,\leftrightarrow \, 4)$ & $0.983$\\[1ex] 
\hline 
\end{tabular}
\label{table:pros-table} 
\end{table}

\section{Simulation Study}
In order to evaluate the accuracy and performance of CDfdr algorithm, we perform two simulated experiments. We are mainly interested to investigate how accurately different methods estimate local fdr specially in the tails. For that purpose, our main criteria will be the MISE, mean integrated square error. Comparisons will be done with Locfdr, Mixfdr and Fdrtool \citep{strimmer}. Grenander density estimation is used in Fdrtool for  estimating the unconditional density $f$, implemented in R package \texttt{Fdrtool}.

\subsection{Mixture Normal.}

We simulate $T_i \sim \cN(\mu_i,1),i=1,\ldots,N=5000$ out of which $4500$ $\mu_i$ is set to zero. The remaining $500$ is drawn from (once and for all) $\cN(\mu,1)$. We estimate the $\widehat \fdr$ for various methods and repeat the whole process $150$ times for $\mu=0.2,0.5,1,2$. Our set up closely follows \cite{storey10} and \cite{Om10}.
\vskip.25em

The goal is to investigate how efficiently the methods can approximate the true fdr when we provide them with the true oracular null density $\cN(0,1)$. The question we asked here: whether there is any extra efficiency gain possible even when we know ``right'' null model. It is not hard to believe that the situation that demand empirically estimating the null will have greater approximation error.
\vskip.25em

Fig 7 depicts the expectation and standard deviation of local fdr for various methods under four different choices of $\mu$. For $\mu=2$, when the signal and noise are well-separated, all of the methods perform equally well apart from Fdrtool, which not only shows high bias but has large variability in the crucial tail region. Under more difficult scenario $\mu=.2$, clearly CDfdr is the only method that can claim to be unbiased. The variability of Mixfdr and CDfdr seems similar though in the extreme (right) tail CDfdr shows more stability.
If we look at the $\rm{Sd}(\widehat \fdr)$ curves for CDfdr and compare with other competing methods, it appears to be the least variable (least dynamic range). Also CDfdr achieves near unbiasedness irrespective of the underlying signal strength, which makes it a reliable tool for large-scale inference problems.
\vskip.25em

It is interesting to examine how all of these methods would perform to estimate the true null proportion under different degree of signal strength. We implemented our Algorithm 1, Minimum Deviance Criteria (MDC) to estimate $\pi_0$. The result is summarized in Fig 8. The simulation was done under the same experimental setup. The boxplot of estimates of $\pi_0$ under different nonnull densities is shown in Fig 8. Locfdr shows largest variability. Mixfdr certainly performs best among the competing methods. However, the method which was particularly successful to estimate the true null proportion $\pi_0=.9$ quickly and accurately is based on the MDC (which is a module of the CDfdr program). This makes the CDfdr algorithm more powerful and efficient.

\begin{figure*}[htb]
\vspace{-1em}
 \centering
 \includegraphics[height=\textheight,width=\textwidth,keepaspectratio]{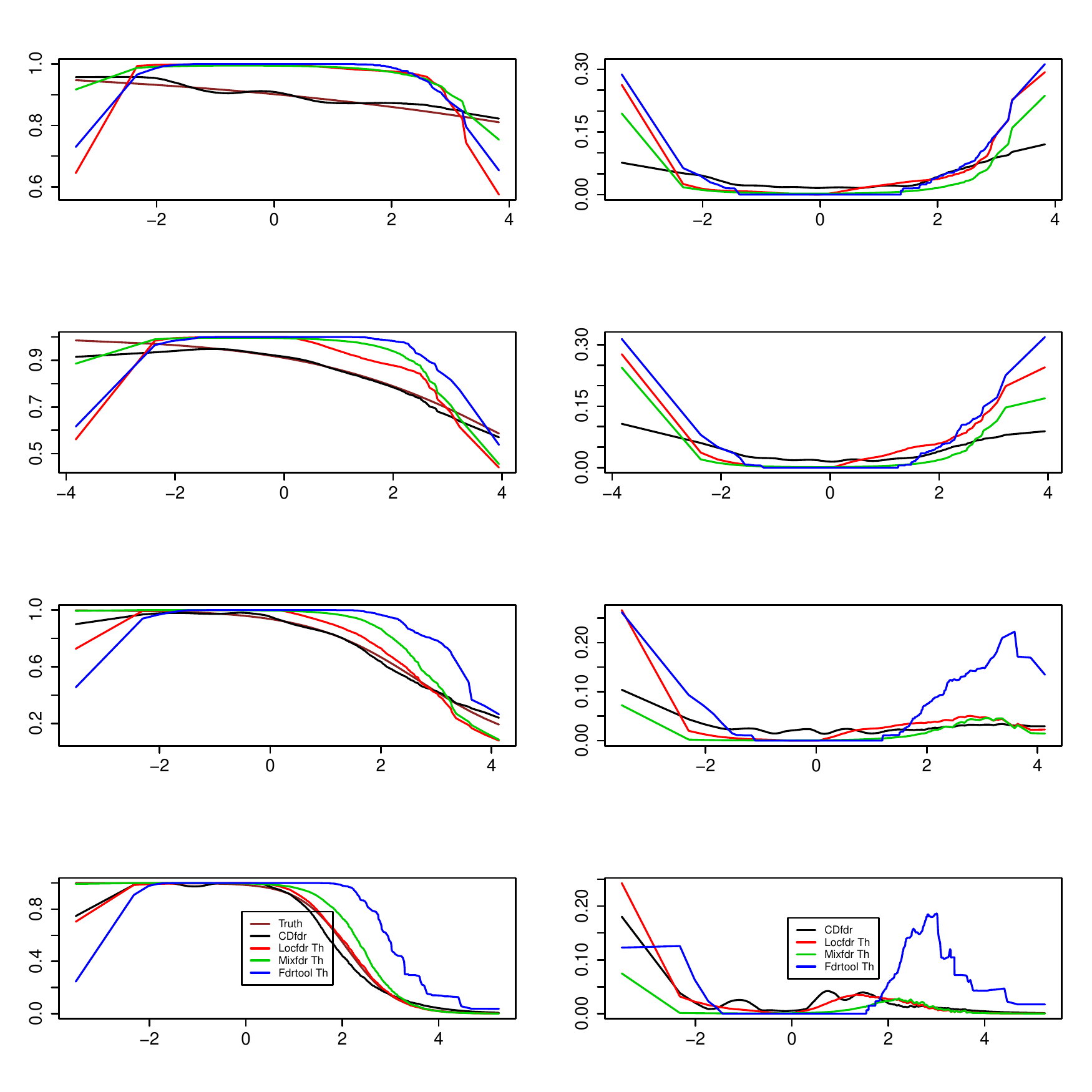} \\
\vspace{-.2em}
\caption{Model $.9 \cN(0,1) + .1 \cN(\mu,1)$. Rows corresponds to $\mu=.2,.5,1,2$ and columns $\Ex(\hat \fdr(z))$ and $\rm{Sd}(\hat \fdr(z))$ for various competing methods.}
\end{figure*}

\begin{figure*}[htb]
\vspace{-1em}
 \centering
 \includegraphics[height=\textheight,width=\textwidth,keepaspectratio]{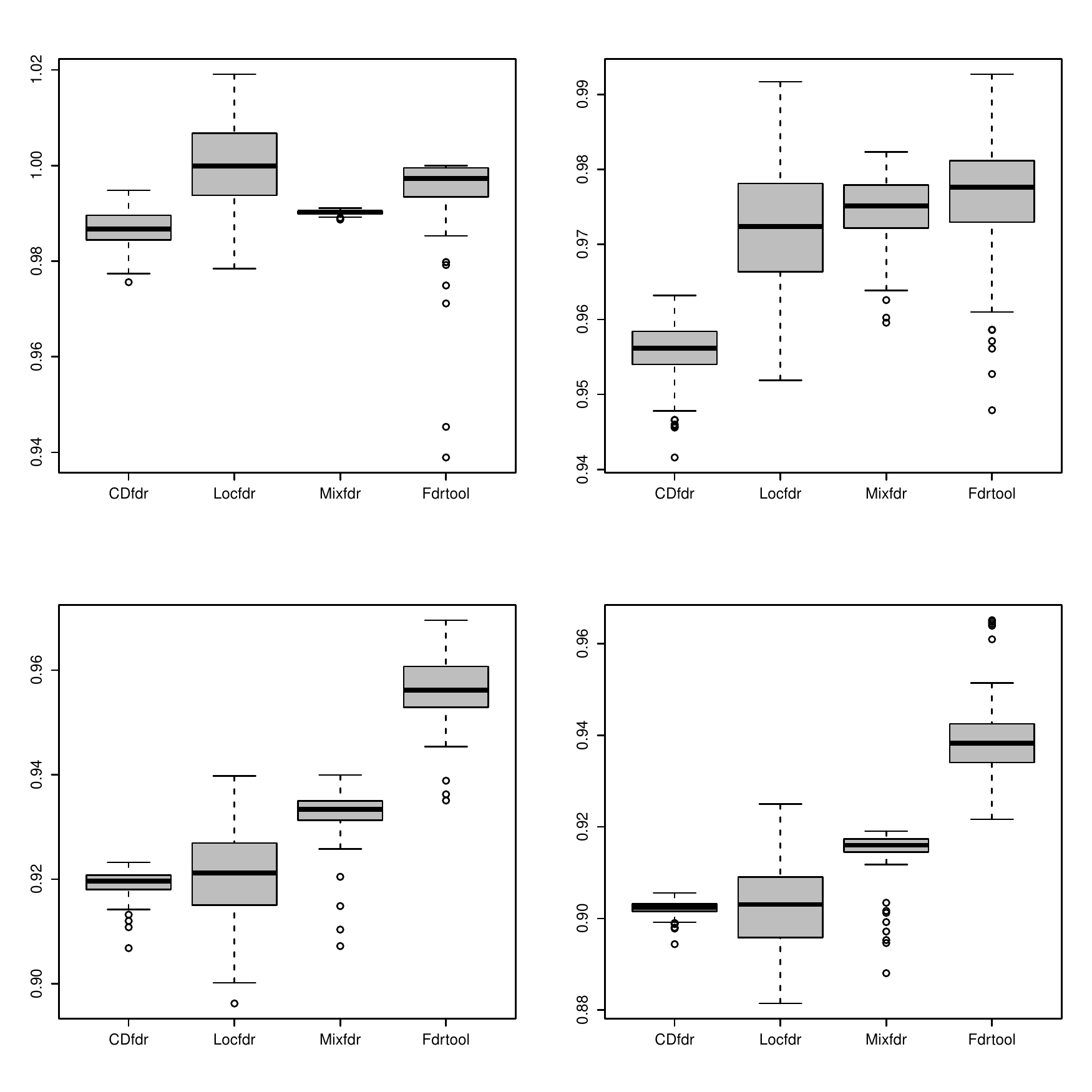} \\
\vspace{-.5em}
\caption{Estimation of $p_0$ for model $.9 \cN(0,1) + .1 \cN(\mu,1)$ under $\mu=1,2,3,4$. It demonstrates how fast and accurately the various methods can learn the parameter $p_0$.}
\end{figure*}

\subsection{Mixture Uniform.}
We generate p-values from the following model:
\beq
\pi_0\, \rm{Uniform}[0,1]\,+\, (1-\pi_0) \, \rm{Uniform}[0,a].
\eeq
We used the following parameter choices: $\pi_0=0.9,0.95,0.99$ and $a=0.02,0.002$. A similar experiment is described in \cite{strimmer,kerfdr}. Here we particularly pay attention to the tails and for what we consider the tail-specific MISE criteria $\E \int_{\mathcal{S}} (\fdr(u) - \hat \fdr(u))^2 \dd u$, where $\mathcal{S}$ denotes the collection of $u$ coming from the alternative model $U[0,a]$. The goal is to quantify how precisely the fdr is estimated for the signals (in the tail). Here the parameter $a$ controls the signal strengths and parameter $\pi_0$ determine the underlying sparsity levels.
\vskip.25em
Our simulation design covers the complete spectrum from
\[ \text{dense and weak}\, \rarrow \, \text{rare and weak} \, \rarrow  \, \text{strong and dense} \, \rarrow \, \text{strong and sparse signal}.  \]
Fig 9 shows the results. In the presence of weak signals ($a=0.02$ first row of Fig 9) the Locfdr and Mixfdr show a great deal of variability. CDfdr maintains the smallest tail-specific MISE among all the methods, which again ensures its utility.  For strong signals (second row of Fig 9), most of the methods have reasonable performance, except perhaps, $(a=0.002,\pi_0=0.9)$ case where the Locfdr poorly approximates the tail. Large number of outliers for Fdrtool are also not satisfactory.
\vskip.25em
Overall, it is encouraging that CDfdr adapts to the underlying signal sparsity and strength in many cases, which makes it very attractive and reliable for large-scale studies. Undoubtedly for the examples we have discussed in this paper it appears, that CDfdr shows the most consistent and robust performance.

\begin{figure*}[htb]
\vspace{-2em}
 \centering
 \includegraphics[height=\textheight,width=\textwidth,keepaspectratio]{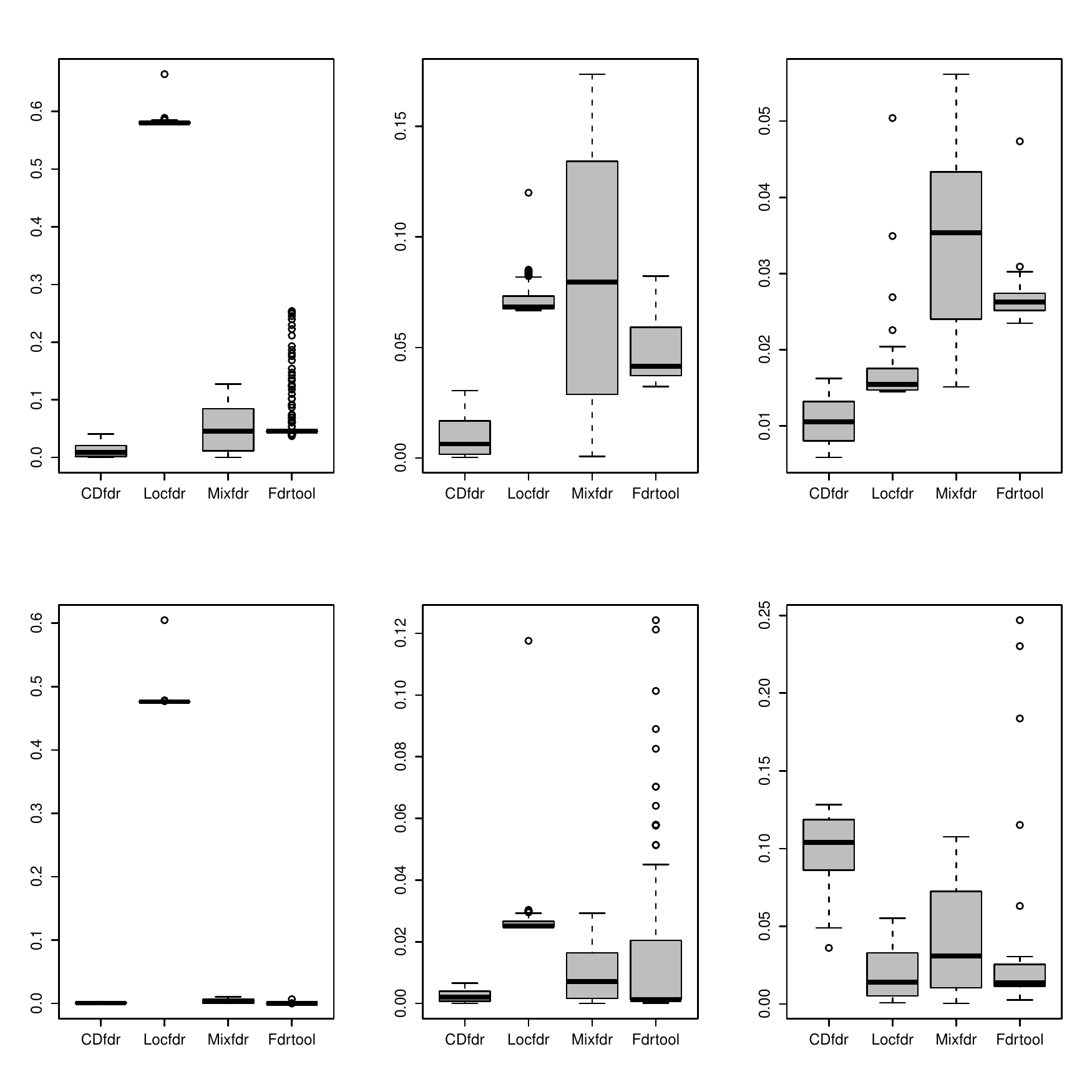} \\
\vspace{-.5em}
\caption{Compares the tail specific MISE for Uniform mixture model $\pi_0\, \rm{Uniform}[0,1]\,+\, (1-\pi_0) \, \rm{Uniform}[0,a]$. The rows corresponds to $a=0.02$ and $0.002$. For each row the columns (from left to right) denotes $\pi_0=0.9,0.95,0.99$.}
\end{figure*}

\section{Related Literature}
In recent years a variety of methods for estimating local false discovery rate has been proposed. Here we review some of the related work.
\vskip.2em
\textbf{I. Parametric Modeling}: The Mixture model is one of the main contenders. The most popular p-value based mixture model is beta-uniform mixture (BUM) introduced by \cite{PM03}, which assumes the following model for comparison density
\beq \label{eq:bum}
d(u;F_0,F)~=~\pi_0 \rm{U}(0,1)\,+\, (1-\pi_0) \rm{Beta}(\al_1,\be_1).
\eeq
This model can be motivated as a quantile domain version of the two-component mixture model proposed by \cite{mc06}, by writing it as
\beq \label{eq:mix}
\dfrac{f(F_0^{-1}(u))}{f_0(F_0^{-1}(u))}~=~\pi_0\,+\,(1-\pi_0) \dfrac{f_1(F_0^{-1}(u))}{f_0(F_0^{-1}(u))},
\eeq
which in our notation can be expressed as,
\beq
d(u;F_0,F)~=~\pi_0 \rm{U}(0,1)\,+\, (1-\pi_0) d(u;F_0,F_1).
\eeq
Under the two-component Gaussian mixture model \citep{mc06}, plot the shape of $d(u;F_0,F_1)$ and compare it with beta density to relate it with Eq. \ref{eq:bum}. A important point to note: the beta-like shape of $d(u;F_0,F_1)$    does not rely on the assumption of Gaussian mixture components. For large class of Gaussian and non-Gaussian density (for both null $f_0$ and non-null $f_1$) it will hold true. In that sense, the BUM model is more general and robust to model mis-specification compared with the Gaussian mixture model formulation.  Extension to more than one mixture components is considered in \cite{all02,Om10} and \cite{Om12}. A regression based exponential comparison density estimator is proposed in \cite{efron08}
\beq
d(u;F_0,F)~=~\exp\{\sum_j \be_j u^j\}.
\eeq
For a comprehensive review refer to \cite*{cheng07}.
\vskip.2em
\textbf{II. Nonparametric Modeling}: In this category, the most popular methods are based on Bernstein polynomials \citep{bern08}, kernel density estimation \citep{kerfdr} and Grenander density \citep{lan05}. For excellent review refer to \cite{strimmer}.
\vskip.3em
All the approaches that we have reviewed here are based on well-known density estimation techniques. The present work proposes an entirely different modeling principle, which has not been considered in the literature.

\section{Concluding Remarks}
We have developed quantile domain machinery that provides a new way of studying large-scale inference problems. The results of the current article have three main contributions:
\vskip.15em
{\bf(a)} they provide a new class of functional statistical inference tools for multiple hypothesis testing problem by using the concept of comparison density.
\vskip.15em
{\bf(b)} they establish connection between local fdr and comparison density based multiple hypothesis testing. This alternative representation ensures direct one-step modeling that might otherwise be difficult to attain, thus addressing the neglected side of false discovery research. This aspect makes it fundamentally different from all the earlier attempts of modeling local fdr.
\vskip.15em
{\bf(c)} they introduce a new density estimation technique based on the idea of pre-flattening smoothing, allowing richer data-driven specification for tail-modeling via simple parametric models, which has an added advantage of being interpretable and easily implementable. Our density estimation procedure (Section 3.3) can be interpreted from at least two angles (algorithmically analogous): semiparametric and (empirical) Bayes. The general density estimation technique that we have provided in Section 3.3 could also be adapted for modeling heavy-tailed data \citep{mark08}.
\vskip.35em
Large-scale inference problems can be studied in two different platforms - distribution domain (original test statistics) or quantile domain (p-values). There are primarily two main reasons why we recommend working with p-values compared with raw test statistics. First, It allows us to estimate the local fdr directly in one-step via comparison density. Second, modeling distribution of $f$ is much more complex than modeling the comparison density using beta-prewhitening transformation. $f$ can potentially take any shape and modality (i.e., number and nature of mixing components) but the density of p-values is guaranteed to have ``U'' like shape for large-scale simultaneous testing problems.
\vskip.15em

There are two broad categories of false discovery methods: distribution function or tail-area based approach  and the density based approach. In this paper we have mainly focused on the density based local fdr methods (CDfdr).
One advantage of the comparison density concept that we have introduced in this work is that it allows us to describe the distribution function based methods as well (CDFdr) using comparison distribution function $D(u)=\int_0^{u} d(u') \dd u'$. An important note: false discovery rate control rule using  Benjamini and Hochberg statistic \citep{BH95}, higher-criticism statistic \citep{HC08}, Berk-Jones statistic \citep{BJ95} and many others can be represented in a unified manner by taking different distance measure between $\widetilde D(u)$ and $u$.

\vskip.15em
It is clear that there are many topics that needs further investigation.
\vskip.15em
{\bf(P1.)} We have established the heuristics and intuition behind comparison density based signal detection in Section 2.1 and provided a justification by connecting it with local fdr technique. There is still the interesting question of whether we can directly establish the validity of the procedure ? It is likely that comparison distribution limit theorems in conjunction with quantile limit theorems \citep{parzen99} can provide some light. Similar empirical process approach has been taken in \cite{wasser04}.
\vskip.15em
{\bf (P2.)} What is the effect of dependence on CDfdr ? The properties are well documented for standard nonparametric density estimators like kernel smoothing \citep{robin83,hart90}. However, systematic theoretical investigation for the proposed pre-flattened density estimator is still unexplored and open problem.
\vskip.15em
{\bf (P3.)} Apart from continuous data, can CDfdr cope with the discrete data ? This is potentially possible if we use mid-probability integral transformation $\Fm(X;X)$ \citep{parzen91b,ma10} and define the concept of mid-pvalue instead of traditional p-value. More research required to design basis functions for estimating discrete $\dhat(u,\Fm_0,F)$ \citep{deepthesis}.
\vskip.15em
{\bf (P4.)} How to extend the concept of CDfdr into two-dimension which is an important practical problem for inferences on images in disciplines like neuroscience and astronomy. Preliminary investigation indicates that the requirement of nonparametric copula density estimator - currently under investigation \citep{D12b,D13a} - could provide an alternative to random-field theory approaches pioneered by \cite{worsley92}.

 We leave answers to all of these questions for further research.

\vskip.15em
 \cite{efron01} proposed empirical Bayes formulation of the frequentist Benjamini and Hochberg's False Discovery Rate method. This article attempts to unify `the two cultures' using concepts of comparison density and distribution function. This work is just a first small step towards large-scale data analysis which enjoys the interplay of frequentist (parametric, semi-parametric, nonparametric) and (empirical) Bayesian methods, facilitated by modern quantile domain tools.

\section*{Acknowledgment} The author would like to thank Professor Emanuel Parzen for several valuable discussions. The author also thanks Editor and anonymous referee for their constructive remarks.
\vskip2em
\bib

\newpage

\end{document}